# Magnetoelectric Properties of Pb Free $Bi_2FeTiO_6$: A Theoretical Investigation

Lokanath Patra[1, 2] and P. Ravindran[1, 2, 3, 4, a)]

[1]*Department of Physics, Central University of Tamil Nadu, Thiruvarur-610005.*
[2]*Simulation Center for Atomic and Nanoscale Materials, Central University of Tamil Nadu, Thiruvarur-610005.*
[3]*Department of Materials Science, Central University of Tamil Nadu, Thiruvarur-610005.*
[4] *Department of Chemistry, Center for Materials Science and Nanotechnology, University of Oslo, P.O. Box 1033 Blindern, N 1035 Oslo, Norway*

a) Corresponding author: raviphy@cutn.ac.in

**Abstract.** The structural, electronic, magnetic and ferroelectric properties of Pb free double perovskite multiferroic $Bi_2FeTiO_6$ are investigated using density functional theory within the general gradient approximation (GGA) method. Our structural optimization using total energy calculations for different potential structures show a minimum energy for a non-centrosymmetric rhombohedral structure with *R3* space group. $Bi_2FeTiO_6$ is found to be an antiferromagnetic insulator with *C*-type magnetic ordering with bandgap value of 0.3 eV. The calculated magnetic moment of 3.52 $\mu_B$ at Fe site shows the high spin arrangement of 3*d* electrons which is also confirmed by our orbital projected density of states analysis. We have analyzed the characteristics of bonding present between the constituents of $Bi_2FeTiO_6$ with the help of calculated partial density of states and Born effective charges. The ground state of the nearest centrosymmetric structure is found to be a G-type antiferromagnet with half metallicity showing that by the application of external electric field we can not only get a polarized state but also change the magnetic ordering and electronic structure in the present compound indicating strong magnetoelectric coupling. The cation sites the coexistence of Bi 6*s* lone pair (bring disproportionate charge distribution) and $Ti^{4+}$ $d^0$ ions which brings covalency produces off-center displacement and favors a non-centrosymmetric ground state and thus ferroelectricity. Our Berry phase calculation gives a polarization of 48 $\mu Ccm^{-2}$ for $Bi_2FeTiO_6$.

## INTRODUCTION

Perovskites based research has gained much attention for last few decades because they show a variety of exotic properties such as superconductivity in $BaBi_{1-x}Pb_xO_3$,[1] colossal magnetoresistance in $La_{1-x}Ca_xMnO_3$,[2] ferroelectricity in $BaTiO_3$[3], multiferroicity in $BiFeO_3$ etc. In case of an ideal $ABO_3$ perovskite, $BO_6$ forms cubic network of corner-sharing octahedra and the *A* cations occupies the 12-coordinate positions between 8 $BO_6$ octahedra. If two different kinds of *B* atoms i.e. *B'* and *B''* are placed on the *B* sublattice, a double perovskite having general formula $A_2B'B''O_6$ is formed. Double perovskites appear promising in the development of spin-polarized conduction and ferromagnetic semiconductors. For example, $La_2NiMnO_6$[4] shows ferromagnetism near room temperature as well as exhibits semiconducting behavior. Bi-based perovskites[5,6] have become more popular due to their high electrical polarization which comes from $Bi^{3+}$ lone pairs. It is also environmentally friendly as $Pb^{2+}$ which has a similar electronic configuration is toxic in nature. Perovskite $BiFeO_3$ is one of the most studied multiferroics, which possess both weak ferromagnetism and ferroelectric characteristics in a single phase. Meanwhile, many have studied the magnetoelectric properties of $BiFeO_3$ experimentally and theoretically. $BiFeO_3$ is known to be ferroelectric with a Curie temperature of about 1103 K and it possesses antiferromagnetic (AFM) ordering with a Néel temperature of 643 K[7,8]. Our previous calculations show that $BiFeO3$ stabilizes with *G*-type AFM ordering. The distorted perovskite structure with rhombohedral symmetry (space group *R3c*) shows a weak ferromagnetism due to the presence of canted spin arrangement. $BaTiO_3$ possesses ferroelectricity which originates from the covalent bonding between non-magnetic $Ti^{4+}$ and $O^{2-}$.[9] Therefore, a new multiferroic double perovskite oxide with high electrical polarization can be designed with $Bi^{3+}$ at the *A* sites and a combination of magnetic as well as non-magnetic (in this case $Ti^{4+}$) ions at *B* sites. Here we report the results for a new multiferroic compound $Bi_2FeTiO_6$ which is designed with the above-mentioned strategy and it is expected to produce better electrical polarization than $BiFeO_3$ because of the coexistence of two mechanisms to bring high polarization. Though there are no experimental data reported for $Bi_2FeTiO_6$, the magnetoelectric behavior is theoretically predicted by Feng *et al.*[10]. But in their study, the electronic structure and magnetic properties are not analyzed in detail which motivated us for the current study.

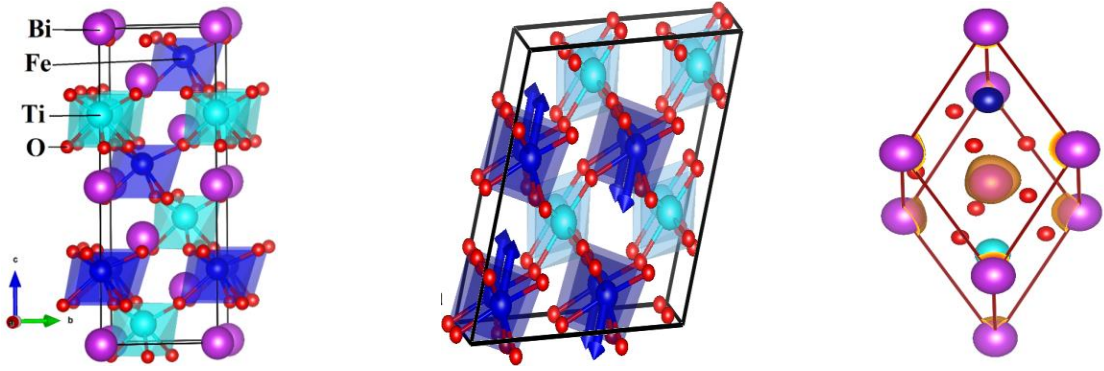

**FIGURE 1.** (color online) (a) The optimized crystal structure of $Bi_2FeTiO_6$ in *R3c* space group, (b) The ground state *C*-type magnetic ordering in a 2x2x2 supercell. (Only Ni-O and Ti-O octahedra are shown for simplicity), (c) The iso-surface plot of charge density with the value of 0.75 e/Å$^3$ for $Bi_2FeTiO_6$. The asymmetric charge distribution around Bi atom shows the presence of lone-pair electrons. (In all the pictures, similar colors are used as denoted in Fig. (a)).

## COMPUTATIONAL DETAILS

The calculations presented here are performed within the framework of density functional theory as implemented in VASP[11] code. We have used GGA with the Perdew-Burke-Ernzerhof (PBE)[12] as exchange and correlation functional. We have shown earlier that the ferroelectric properties are very sensitive to structural parameters[5] and hence we have used a very high energy cut-off of 800 eV. A Monkhorst pack[13] 6x6x6 **k**-point mesh is used for the ferromagnetic calculations and similar density of **k**-points are used for all other calculations. In order to identify the magnetic ground state, we have considered possible collinear magnetic orderings i.e. ferromagnetic (FM), *A*-type antiferromagnetic (*A*-AFM), *C*-type antiferromagnetic (*C*-AFM), and *G*-type antiferromagnetic (*G*-AFM) orderings[14]. The convergence criteria were set to $10^{-6}$ eV per cell in the case of total energy and 1 meVÅ$^{-1}$ in case of the forces acting on the atoms. The Born-effective charges (BEC) were calculated using the so-called "Berry phase finite difference approach"[15].

## RESULTS AND DISCUSSIONS

The ground state structure for $Bi_2FeTiO_6$ was obtained by optimizing the geometries of a range of starting configurations: cubic $Fm\overline{3}m$, tetragonal *P4mm*, tetragonal *Amm2*, rhombohedral *R3c*, rhombohedral $R\overline{3}c$, monoclinic *C121*, and monoclinic *C2/c*. Using force as well as stress minimization, the structural relaxations were performed for different volumes. The non-centrosymmetric structure with *R3c* symmetry was found to have the lowest energy among all the considered structural configurations. The optimized equilibrium volume was found to be 127.82 Å$^3$ and the lattice parameter and the rhombohedral angle were calculated to be 5.62 Å and 60.77°, respectively. The force minimized atomic positions for the ground state configurations are: Bi (1a) (0.0, 0.0, 0.502), Fe/Ti (1a) (0.5, 0.5, 0.234/0.728), O1 (3a) (0.563, 0.928, 0.388), and O2 (3b) (0.882, 0.460, 0.038). The optimized ground state crystal structure for $Bi_2FeTiO_6$ is given in Fig. 1(a). It can be seen from the figure that Fe and Ti form corner shared distorted octahedra with the neighboring O ions.

Our total energy calculations predict a ground state *C*-AFM ordering (see Fig. 1(b)) for $Bi_2FeTiO_6$. The *C*-AFM state is lower in energy than FM, *A*-AFM and *G*-AFM states by 77 meV, 25 meV, and 5 meV respectively. The presence of lone pair electrons at Bi site (see Fig. 1(c)) and the Ti ions with $d^0$ configuration implies that Fe tends to have an oxidation state of +2. In an ideal octahedral cubic crystal field, the *d*-level splits into triply degenerate $t_{2g}(d_{xy}, d_{xz}, d_{yz})$ and doubly degenerate $e_g(d_{x^2-y^2}, d_{z^2})$ levels. If the system is purely ionic, $Fe^{2+}$ with 6 *d* electrons will fill these energy levels with a low spin state ($t_{2g}^6$, $d_{x^2-y^2}^0$, $d_{z^2}^0$) or intermediate spin state ($d_{xy}^2$, $d_{xz}^2$, $d_{yz}^1$, $d_{x^2-y^2}^1$, $d_{z^2}^0$) or high spin state ($d_{xy}^2$, $d_{xz}^1$, $d_{yz}^1$, $d_{x^2-y^2}^1$, $d_{z^2}^1$) with spin moments 0 $\mu_B$, 2 $\mu_B$, and 4 $\mu_B$ respectively. In this case, the $Fe^{2+}$ can be assigned a HS state as the magnetic moment at the Fe site is found to be 3.52 $\mu_B$. Due to the presence of covalent interaction of $Fe^{2+}$ with neighboring oxygens, we get a reduced non-integer value of spin moment than that of the pure ionic case. It may be noted that due to this covalent interaction, one can get induced magnetic moment at O site. Consistent to this view point, our calculation shows the magnetic moment at O sites varies between 0.1 $\mu_B$ to 0.5 $\mu_B$. It can be seen from our orbital projected DOS given in Fig. 2(a) that, all the 5 *d* orbitals of $Fe^{2+}$ ions are occupied in the majority spin channel with 5 electrons. The remaining one electron is mainly distributed in the minority spin channel of $d_{xy}$, $d_{xz}$, $d_{yz}$, and $d_{x^2-y^2}$ orbitals indicating the importance of covalency effect to understand the electron occupation in various orbitals. We have calculated the occupation number for these orbitals and are: $d_{xy}$ = 0.29, $d_{xz}$ = 0.27, $d_{yz}$ = 0.23, and $d_{x^2-y^2}$ = 0.17; the sum of electron occupation in these four orbitals is ~ 1. It can also be seen that, the $t_{2g}$ states of Ti *d* are nearly empty and finite DOS can be seen at $e_g$ states due to covalency effect. These results confirm that Ti is in a 4+ state in $Bi_2FeTiO_6$. We also found that, $Bi_2FeTiO_6$ stabilizes with *G*-AFM ordering in the centrosymmetric $R\bar{3}c$ structure with half metallicity.

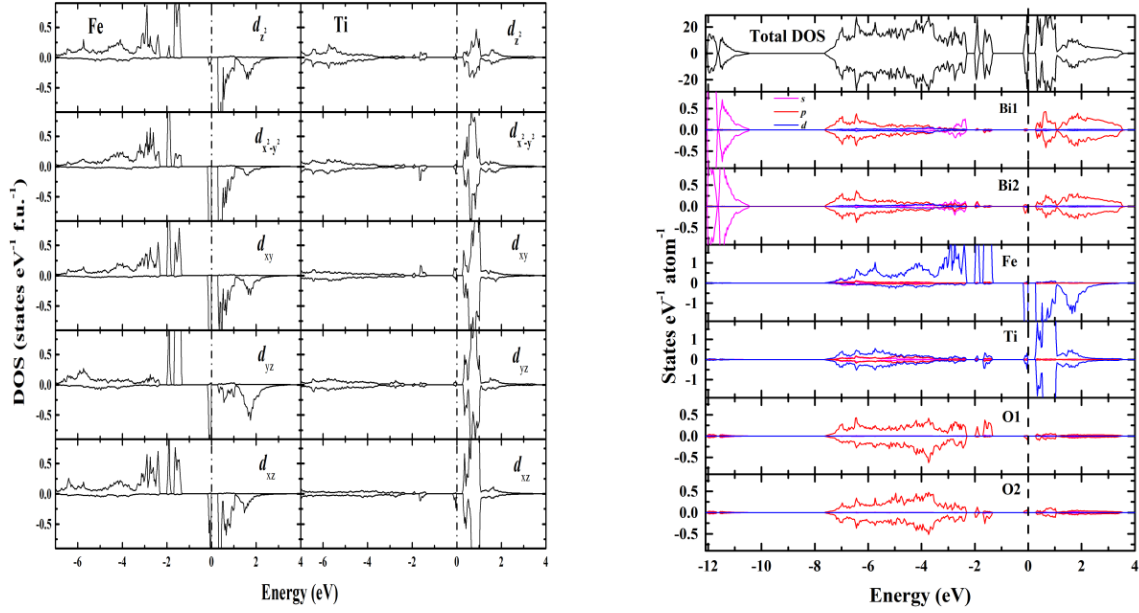

**FIGURE 2.** (color online) (a) The calculated orbital-projected density of states of $Fe^{2+}$ and $Ti^{4+}$ for $Bi_2MnTiO_6$ in the ground state *C*-AFM ordering. (b) The total and partial density of state plots for $Bi_2MnTiO_6$ in the ground state *C*-AFM ordering.

The total and partial density of states (DOS) for $Bi_2FeTiO_6$ in the ground state *C*-AFM ordering are shown in Fig. 2(b). The calculated bandgap is ~ 0.3 eV showing semiconducting behavior. The DOS curves show that the Bi 6*p*, Ni 3*d*, Ti 3*d* and O 2*p* states are spreading over an energy range of -8 eV to Fermi level ($E_F$). This ensures the presence of covalent interaction among the constituents as the presence of degenerated electronic energy states is an essential condition for covalency. Due to this covalent interaction between Ti 3*d* and O 2*p* states, the charges redistribute between these states resulting in the occupation of Ti 3*d* states as evident from Fig. 2(a) and Fig. 2(b). The Bi 6*s* lone pairs are well localized around -11 eV which can be seen from the DOS curve in Fig. 2(b). Interestingly, consistent with previous observation[10], we have also found an insulating state for the FM configuration though the FM state is 77 meV/f.u. higher than the *C*-AFM state. It may be noted that the ferromagnetism and insulating behaviors seldom coexist and such ferromagnetic insulators have a wide range of applications in spintronics field.

To study the ferroelectric properties of $Bi_2FeTiO_6$, we have calculated the Born effective charge (BEC, Z*) with the Berry phase method generalized to the spin-polarized system. The calculated average values of the BEC are: $Z^*_{Bi1}$ = 5.07, $Z^*_{Bi2}$ = 4.82, $Z^*_{Fe}$ = 2.15, $Z^*_{Ti}$ = 6.05, $Z^*_{O1}$ = -3.01, $Z^*_{O2}$ = -3.02. Our calculations show finite values of off-diagonal elements in the BEC tensor for $Bi_2FeTiO_6$ which confirms the covalent interaction between O and transition metals. BECs can also be used to quantify the spontaneous electrical polarization in $Bi_2FeTiO_6$. As the paraelectric phases lack off-center displacement, they produce zero net polarization. $Bi_2FeTiO_6$ contains distorted octahedra in which the atoms are shifted from the nearest centrosymmetric space group $R\bar{3}c$. The calculated polarization $Bi_2FeTiO_6$ with present Berry phase method is ~ 48 μCcm$^{-2}$. Our partial polarization analysis shows that the Ti $3d$ – O $2p$ covalence interaction contribute more to the net polarization than the $Bi^{3+}$ lone pair electrons. The value of polarization has not come up to our expectations based on the coexistence of two ferroelectricity mechanism in this compound. This lower value of polarization than that of $BiFeO_3$ can be partly understood by the smaller BECs for Bi, Fe and O in $Bi_2FeTiO_6$ than those in $BiFeO_3$. It can also be seen that Ti has a large BEC i.e. +6.05 as compared to its nominal charge of +4 showing the substantial contribution coming from dynamical charges associated with covalency which dominate the Ti off-center displacement. The addition of Ti in $BiFeO_3$ has not only suppressed the ferroelectric properties, but also has affected the magnetic properties by changing the oxidation state of Fe from 3+ state to 2+ state. As a result, the obtained magnetic moment at Fe site in $Bi_2FeTiO_6$ is found to be 3.52 $\mu_B$ compared to a value of 3.79 $\mu_B$ in the Fe site in $BiFeO_3$. It may also be noted that the applied electric field can not only polarize the centrosymmetric structure but also it can change the magnetic ordering from $G$-type to $C$-type along with half metallic state to an insulating state. We have also calculated polarization using the nominal charges which produced a polarization of ~ 18 μCcm$^{-2}$ which is almost 62% less than the polarization values calculated using BECs. This reduction in polarization is due to the strong covalency present in the system. Consequently, our primary aim to design a magnetoelectric material is fulfilled as both magnetism and ferroelectricity coexist in $Bi_2FeTiO_6$.

## CONCLUSION

$Bi_2FeTiO_6$ is predicted to be a magnetoelectric multiferroic with $C$-type magnetic ordering. The $Fe^{3+}$ ions are found to be in an HS state with a spin moment of 3.52 $\mu_B$. The presence of lone pair at Bi site which is confirmed from our isosurface charge density plot and the strong covalent bonding between Ti $d$ and O $p$ states which is evident from the partial DOS and BECs analyses, are the primary reasons for the stabilization of non-centrosymmetric structure over the centrosymmetric structure. Our Berry phase calculation produced a spontaneous electrical polarization of 48 μCcm$^{-2}$. The insulating ferromagnetic state of $Bi_2FeTiO_6$ needs more attention for applications in the field of spintronics. Our study suggests that double perovskites having formula $Bi_2MM^{/}O_6$ (M and M$^{/}$ are combinations of magnetic – non-magnetic or magnetic – magnetic atoms) can be good candidates for multiferroic applications. We have also shown that by applying electric field one can change the magnetic ordering from C-AFM to $G$-AFM state and change the electronic state from insulator to half metal in $Bi_2FeTiO_6$.

## ACKNOWLEDGMENTS

The authors are grateful to the Research Council of Norway for providing computing time at the Norwegian supercomputer facilities. This research was supported by the Indo-Norwegian Cooperative Program (INCP) via Grant No. F. No. 58-12/2014(IC)